\documentclass[12pt]{article}
\usepackage{psfig,epsfig,bezier}

\newcommand{\be}{\begin{equation}}
\newcommand{\ee}{\end{equation}}
\newcommand{\ba}{\begin{eqnarray}}
\newcommand{\ea}{\end{eqnarray}}
\begin {document}
\def\chic#1{{\scriptscriptstyle #1}}
\begin{flushright}
{\bf hep-ph/0211437}\\
\end{flushright}
\begin{center}
{\Large \bf One-Loop Electroweak Corrections to the Muon Anomalous
Magnetic Moment Using the Pinch Technique}\\
\vspace{1.3cm}

{\large L. G. Cabral-Rosetti$^{a ,}$\footnote{E-mail: luis@nuclecu.unam.mx}
, G. L\'opez Castro$^{b ,}$\footnote{E-mail: Gabriel.Lopez@fis.cinvestav.mx}
and J.Pestieau$^{c ,}$\footnote{E-mail: pestieau@fyma.ucl.ac.be}
}\\   

\

{\small \it $^a$Instituto de Ciencias Nucleares, Universidad Nacional
Aut\'onoma de M\'exico, \\ Circuito Exterior C.U., A.P. 70-543,
04510 M\'exico D.F. M\'exico 
\\
$^b$Departamento de F\'{\i}sica, Centro de Investigaci\'on y de Estudios
\\ Avanzados del IPN, Apartado Postal 14-740, 07000 M\'exico D.F. M\'exico
\\
$^c$ Institut de Physique Th\'eorique, Universit\'e catholique \\
de Louvain, B-1348 Louvain-La-Neuve, Belgium}
\end{center}
\vspace{1.3cm}

\begin{abstract}
The definition of the physical properties of particles in perturbative
gauge theories must satisfy gauge invariance as a requisite. The Pinch
Technique provides a framework to define the electromagnetic form factors
and the electromagnetic static properties of fundamental particles in a
consistent and gauge-invariant form. We apply a
simple prescription derived in this formalism to check the calculation of
the gauge-invariant one-loop bosonic electroweak corrections to the muon
anomalous magnetic moment.
\end{abstract}

\

\

A definition of the neutrino charge radius that satisfies good physical
requirements, i.e. it is a physical observable, has been provided recently
\cite{ncr} in the framework of the Pinch Technique (PT) formalism \cite{pt}. 
In the PT formalism, the construction of a gauge-independent and 
gauge-invariant one-loop vertex and, in particular, of an {\it effective} 
electromagnetic form factor for the neutrino amounts to compute \cite{ncr} 
the one-loop vertex corrections using a simple prescription in the linear 
$R_\xi^L$ gauge, where gauge-boson propagators 
\begin{equation}
P_{{\mu}{\nu}}^{\chic V} (q) = {{-\ i}\over {q^{\chic 2} -
M_{\chic V}^{\chic 2}}} \Bigg[ g_{{\mu}{\nu}} + (1-\xi) 
{{q_{\mu} q_{\nu}}\over {\xi q^{\chic 2} - M_{\chic V}^{\chic 2}}} 
\Bigg ]\ ,
\end{equation}
are taken in the  't Hooft-Feynman gauge $\xi=1$, and the usual
three-boson vertex
\begin{equation}
\Gamma_{\alpha \mu \nu}(q,k,-q-k)=(q-k)_{\nu}g_{\alpha \mu} +
(2k+q)_{\alpha}g_{ \mu \nu}-(2q+k)_{\mu}g_{\alpha \nu}\ ,
\end{equation}
is replaced by the truncated vertex \cite{descomp}:
\begin{equation}
\Gamma^F_{\alpha \mu \nu} =(2k+q)_{\alpha}g_{ \mu \nu}+2q_{\nu}g_{\alpha
\mu}-2q_{\mu}g_{\alpha \nu}\ ,
\end{equation}
which satisfies \cite{ncr} a simple Ward identity:
\[
q^{\alpha}\Gamma^F_{\alpha \mu \nu}=(k+q)^2g_{\mu \nu} -k^2g_{\mu \nu}\ .
\]

 We emphasize that this prescription should be applicable not only for the
case of the neutrino but also for the electromagnetic form factors of
quarks and leptons \cite{ql}. In particular, it looks very appealing to
compute in a simple
form the electroweak  contributions to the static properties of fermions.
In this note we apply the PT prescription to give an alternative
evaluation of the one-loop $W$-boson contribution to the anomalous
magnetic moment of the muon, $a_{\mu}\equiv (g-2)/2$. 

   The complete one-loop electroweak corrections to $a_{\mu}$ were 
computed long time ago in refs. \cite{aweak1} (the Higgs boson
contribution and subleading muon mass terms are neglected):
\begin{equation}
a_{\mu}^{weak} =\frac{G_Fm_{\mu}^2}{8\pi^2\sqrt{2}} \left \{ \frac{10}{3}
+\frac{1}{3} [(1-4\sin^2_{\theta_W})^2-5] \right\} \ .
\end{equation}
The first term in Eq. (4) accounts for the $W$-boson (plus
unphysical scalars) contributions and the
second term for the $Z^0$-boson correction to the vertex. Each one of
these contributions is independent of the $\xi$-gauge parameters (in the
linear $R_\xi^L$ gauges) \cite{aweak1}. In the
following, we are concerned first with the derivation of the first term in
Eq. (4) using the PT prescription mentioned above. It is worth to mention 
that, in contradistinction with the Pinch Technique,  the
evaluation of the muon anomalous magnetic form factor (for a non-vanishing
$q^2$ value) is gauge-dependent with the methods used in refs.
\cite{aweak1}.

\begin{figure}[htb]
\centerline{
\epsfig{figure=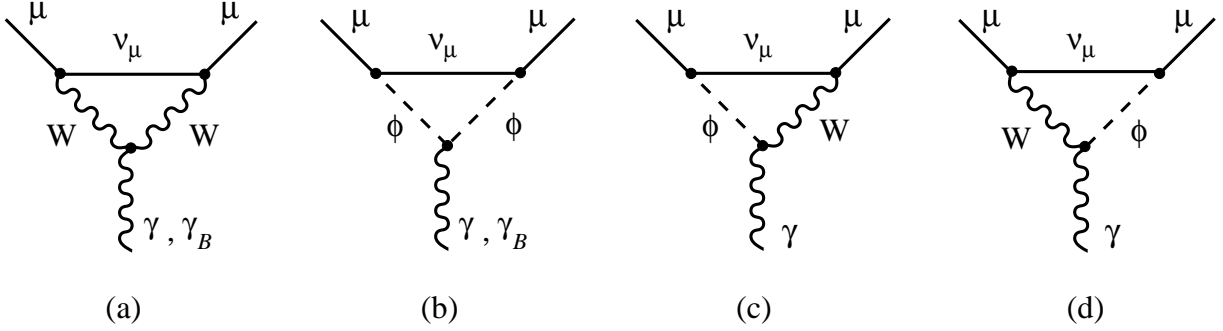,height=4.5cm}}
\caption{$W$-boson contributions to $a_{\mu}$: in the BFM formalism 
only diagrams (a-b) are required, while the four diagrams are necessary
in the $R_\xi^L$ gauge calculation.}
\end{figure}

  Instead of performing an explicit evaluation of the $W$-boson
corrections to the vertex, we can take advantage of a result derived, in
another context, by Brodsky and
Sullivan and independently by Burnett and Levine  in the late
sixties \cite{bs}. Using the $W$-boson propagator of Eq. (1)  
and the electromagnetic vertex of the $W$-boson as proposed by Lee and 
Yang \cite{leeyang} (all particles are
incoming, namely $k_1+k_2+k_3=0$):
\begin{equation}
V_{\mu \alpha \beta}= ie \{ g_{\alpha \beta}(k_1-k_2)_{\mu} -g_{\alpha
\mu} (k_1+\kappa_W k_1+\xi k_2+\kappa_W k_2)_{\beta} 
 +g_{\beta \mu} (k_2+\kappa_W k_2+\xi k_1+\kappa_W
k_1)_{\alpha}\ \}\ ,
\end{equation}

\noindent
it can be shown that the prescription of the PT formalism for the
$W$-boson propagator and electromagnetic vertex (see eqs. (1) and (3)) is
obtained by
choosing\footnote{The usual electromagnetic vertex 
for the $W$-boson in gauge theories is recovered for the special choice
$\xi=0$ and $\kappa_W=1$ in eq. (5).}:
\begin{equation}
\xi=1\ \ \mbox{\rm and} \ \ \kappa_W=1\ . 
\end{equation}

  The $W$-boson contribution (Fig. 1a) to $a^{weak}_{\mu}$  obtained in
refs.
\cite{bs} using
the Feynman rules of Eqs. (1) and (5) is:
\begin{equation}
a_{\mu}^{WW}=\frac{G_F m_{\mu}^2}{8\pi^2\sqrt{2}} \left\{ 2(1-\kappa_W)\ln
\xi + \frac{10}{3}  \right\}\ .
\end{equation}
As it can be easily checked by inserting the values given in eq. (6), the
PT prescription for this correction gives the correct result for the
$W$-boson contributions to $a_{\mu}$ (first term in Eq. (4)). The
contribution from the $Z^0$-boson corresponding to the PT prescription 
($\xi=1$) computed in \cite{aweak1} must be added to Eq. (7). Therefore,
we recover, in the leading muon mass approximation,  the usual result for
the electroweak corrections to $a_{\mu}$ at the one-loop level. 

  As a further check of the results obtained using the PT formalism, we
have computed the one-loop electroweak corrections to $a_{\mu}$ using the
Background Field Method (BFM)
\cite{bfm}. It has been shown that when we fix the gauge parameter in the
electroweak BFM to a particular value, namely $\xi_Q=1$, one recovers the
results of the PT formalism at the one-loop level \cite{bfmxi1}. This
equivalence of both formalisms has been shown to hold at the two-loop
level in Ref. \cite{bfmxi2} and very recently it has been proved at all
orders in \cite{bfmxi3}. Since
the different diagrams that contribute to the S-matrix amplitude in the
framework of the PT have been rearranged in a gauge-invariant form to
produce an effective electromagnetic vertex, this becomes a useful 
test of the calculation. Using the BFM formalism, the calculation of the
$W$-boson and two-scalar contribution (The BFM only requires the
contribution of  
Figs.~1a and 1b) to the anomalous magnetic moment of the muon  becomes (we
keep some terms of higher order in the muon mass):
\begin{eqnarray}
a_{\mu}^{\chic W} \big|_{\chic {Fig. 1a}}^{\chic {BFM}}
&=&\frac{G_Fm_{\mu}^2}{8\pi^2\sqrt{2}} \left\{ \frac{10}{3} +
\frac{5}{6} \left(\frac{m_{\mu}}{m_W} \right)^2 + \frac{7}{6}
\left(\frac{m_{\mu}}{m_W} \right)^4 +\cdots \right\}\ ,  \\
a_{\mu}^{\chic {\phi \phi}} \big|_{\chic {Fig. 1b}}^{\chic {BFM}}
&=& \frac{G_Fm_{\mu}^2}{8\pi^2\sqrt{2}} \left\{
- \frac{1}{3} \left(\frac{m_{\mu}}{m_W} \right)^2 - \frac{1}{4}
\left(\frac{m_{\mu}}{m_W} \right)^4 +\cdots \right\}\ .
\end{eqnarray}
The sum of Eqs. (8) and (9) give :
\begin{equation}
a_{\mu} \big|_{\chic {Fig. 1a+b}}^{\chic {BFM}}
=\frac{G_Fm_{\mu}^2}{8\pi^2\sqrt{2}} \left\{ \frac{10}{3} +
\frac{1}{2} \left(\frac{m_{\mu}}{m_W} \right)^2 + \frac{11}{12}
\left(\frac{m_{\mu}}{m_W} \right)^4  +\cdots \right\}\ ,
\end{equation}
in good agreement with the results obtained using the PT recipe and
with previous results \cite{aweak1}. As a
further check of the subleading terms in $m_{\mu}$, one can compute the 
different contributions from Figs. (1.a-d) in the linear $R_\xi^L$ gauge.
 Choosing the gauge $\xi=1$ we obtain: 

\begin{eqnarray}
a_{\mu}^{\chic W} \Big|_{\chic {Fig. 1a}}^{\chic {R_{\xi}^L}}
&=&\frac{G_Fm_{\mu}^2}{8\pi^2\sqrt{2}} \left\{ \frac{7}{3} +
\frac{1}{2} \left(\frac{m_{\mu}}{m_W} \right)^2 + \frac{5}{6}
\left(\frac{m_{\mu}}{m_W} \right)^4 +\cdots \right\}\ , \\
a_{\mu}^{\chic {\phi \phi}} \Big|_{\chic {Fig. 1b}}^{\chic {R_{\xi}^L}}
&=& \frac{G_Fm_{\mu}^2}{8\pi^2\sqrt{2}} \left\{
- \frac{1}{3} \left(\frac{m_{\mu}}{m_W} \right)^2 - \frac{1}{4} 
\left(\frac{m_{\mu}}{m_W} \right)^4 +\cdots \right\}\ , \\
a_{\mu}^{\chic W} \Big|_{\chic {Fig. 1c}}^{\chic {R_{\xi}^L}}
= a_{\mu}^{\chic W} \Big|_{\chic {Fig. 1d}}^{\chic {R_{\xi}^L}}
&=& \frac{G_Fm_{\mu}^2}{8\pi^2\sqrt{2}} \left\{ \frac{1}{2} +
\frac{1}{6} \left(\frac{m_{\mu}}{m_W} \right)^2 + \frac{1}{6}
\left(\frac{m_{\mu}}{m_W} \right)^4 +\cdots \right\}\ .
\end{eqnarray}
When we add Eqs. (11--13), we obtain the same result as in Eq. (10). Let 
us note that to
obtain the above results we have made extensive use of the expressions
for the two- ($B_0$) and three-point ($C_0$) Passarino-Veltman functions 
derived in Ref. \cite{luis}.

In summary, the application of the prescription given in
Eqs. (1) (with $\xi=1$) and (3), shows the robustness and simplicity of
the PT formalism.
In particular, the PT could be useful to verify the
independence of the result with respect to the gauge-parameter in a
given gauge structure and to clarify the evaluation
of the complete contributions to the
two-loop electroweak corrections to $a_{\mu}$, since it has been proved
that gauge invariance is satisfied to all orders
\cite{bfmxi2,bfmxi3} in this method. Note that the two-loop electroweak
contributions to $a_{\mu}$ were computed in ref.
\cite{czar-mar}. These corrections were computed using the
linear $R_{\xi}$ gauge in the 't Hooft-Feynman gauge and also a nonlinear
gauge structure, and 
neglecting the contributions that involve two or more scalar couplings
\cite{czar-mar} since they are supressed by additional powers of
$m_{\mu}^2/m_W^2$. The two-loop electroweeak corrections amount to a
reduction of --22.6\% with respect to the one-loop electroweak result and 
it is at the level of the sentitivies expected in current experiments.
The PT formalism can therefore provide an  additional check of these
results in a consistent, gauge-invariant and gauge-parameter
independent way.

\

{\bf Acknowledgements}: L. G. C. R. has been partially supported  by
grants No. {\tt IN109001} (DGAPA-UNAM ) and  {\tt I37307-E} (Conacyt).
G.L.C. acknowledges partial financial support from Conacyt .


\begin{thebibliography}{40}
%
\bibitem{ncr} J. Bernabeu, J. Papavassiliou and J. Vidal, Phys. Rev. Lett.
{\bf 89}, 101802 (2002); Erratum Phys. Rev. Lett. {\bf 89}, 229902-1
(2002) and eprint hep-ph/0210055; J. Bernabeu, L. G.
Cabral-Rosetti, J. Papavassiliou and J. Vidal,
Phys. Rev. {\bf D62}, 113012 (2000); L. G. Cabral-Rosetti, Ph. D. Thesis,
Univ.  of Valencia (2000).
%
\bibitem{pt} J. M. Cornwall and J. Papavassiliou, Phys. Rev. {\bf D40},
3474 (1989); J. Papavassiliou, Phys. Rev. {\bf D41}, 3179 (1990).
%
\bibitem{descomp}G.~'t Hooft, Nucl. Phys.{\bf B33}, 173 (1971); J. M. 
Cornwall and G. Tiktopoulos, Phys. Rev. {\bf D15}, 2937 (1977).

\bibitem{ql} J. Papavassiliou and C. Parrinello, Phys. Rev. {\bf D50},
3059 (1994); J. Papavassiliou and A. Pilaftsis, Phys. Rev. Lett. {\bf 75},
3060 (1995); Phys. Rev. {\bf D53}, 2128 (1996).
%
\bibitem{aweak1} K. Fujikawa, B. W. Lee and A. I. Sanda, Phys. Rev. {\bf
D6}, 2923 (1972); I. Bars and M. Yoshimura, Phys. Rev. {\bf D6}, 374
(1972); W. A. Bardeen, R. Gastmans and B. E. Lautrup, Nucl. Phys. {\bf
B46}, 315 (1972); R. Jackiw and S. Weinberg, Phys. Rev. {\bf D5}, 2396
(1972). 
%
\bibitem{bs}
S. J. Brodsky and J. D. Sullivan, Phys. Rev. {\bf 156}, 1644 (1967); T.
Burnett and M. J. Levine, Phys. Lett. {\bf B24}, 467 (1967).
%
\bibitem{leeyang} 
T. D. Lee and C. N. Yang, Phys. Rev. {\bf 128}, 885 (1962).
%
\bibitem{bfm}
A. Denner, G. Weiglein and S. Dittmaier, Nucl. Phys. {\bf B 440}, 95 (1995)
%
\bibitem{bfmxi1}
A. Denner, G. Weiglein and S. Dittmaier Phys. Lett. B {\bf 33}, 420
(1994); S. Hashimoto, J. Kodaira, Y. Yasui and K. Sasaki, 
Phys. Rev. D {\bf 50}, 7066 (1994); J. Papavassiliou, Phys. Rev. D {\bf
51}, 856 (1995).
\bibitem{bfmxi2}
J. Papavassiliou, Phys. Rev. Lett. {\bf 84}, 2782 (2000); D. Binosi and
J. Papavassiliou, Phys. Rev. D {\bf 65}, 085003 (2002); D. Binosi and
J. Papavassiliou, Phys. Rev. D {\bf 66}, 076010 (2002).
%
\bibitem{bfmxi3}
D. Binosi and J. Papavassiliou, e-Print Archive: {\tt hep-ph/0208189}.


\bibitem{luis} L. G. Cabral-Rosetti and M.A. Sanchis-Lozano J. Comput.
Appl. Math. {\bf 115} 93-99 (2000); L. G. Cabral-Rosetti, J. Bernabeu, J.
Vidal, A. Zepeda, Eur. Phys. J. C {\bf 12} 633-642 (2000); L.
G. Cabral-Rosetti and M.A. Sanchis-Lozano Proceedings of {\it 2nd
International Workshop on Graphs, Operads, Logic, Paralel Computation
Mathematical  Physics}, Cuautitlan, Mexico, 6-16 May 2002. e-Print
Archive: {\tt  hep-ph/0206081}.
\bibitem{czar-mar}
A. Czarnecki, B. Krause and W. J. Marciano, Phys. Rev. {\bf D52}, R2619
(1995); Phys. Rev. Lett. {\bf 76}, 3267 (1996).
\end{thebibliography}
\end{document}